\documentclass[lettersize,journal]{IEEEtran}

\usepackage{amsmath,amsfonts}
\usepackage{algorithmic}
\usepackage{algorithm}
\usepackage{array}
\usepackage{textcomp}
\usepackage{stfloats}
\usepackage{url}
\usepackage{verbatim}
\usepackage{graphicx}
\usepackage{cite}
\usepackage[table,xcdraw]{xcolor}
\usepackage{colortbl}
\usepackage{hyperref}

\usepackage{subcaption}

\begin{document}

\title{The energetic cost of mitigating AI attacks in cellular networks}

\author{Adrián Losada, Hao Qiang Luo-Chen, David Segura, Carlos S. Alvarez-Merino, Milan Groshev,   
Emil J. Khatib \IEEEmembership{Member, IEEE}, and Raquel Barco

\thanks{Hao Qiang Luo-Chen, David Segura, Carlos S. Álvarez-Merino, Emil J. Khatib and Raquel Barco are with Telecommunication Research Institute (TELMA), Universidad de Málaga, E.T.S. Ingeniería de Telecomunicación, Bulevar Louis Pasteur 35, 29010, Málaga (Spain). Corresponding author: Hao Qiang Luo-Chen (e-mail: hao@ic.uma.es).}

\thanks{Milan Groshev is with The Laude Technology Company, S.L. 28037 Madrid, Spain. Adrián Losada was with Laude when this work was carried out.}
}

\markboth{Journal of \LaTeX\ Class Files,~Vol.~14, No.~8, August~2021}%
{Shell \MakeLowercase{\textit{et al.}}: A Sample Article Using IEEEtran.cls for IEEE Journals}


\maketitle

\begin{abstract}

The integration of Artificial Intelligence (AI), generally as Machine Learning (ML) algorithms, in all levels and aspects of cellular networks demonstrates the success of data-driven algorithms; for example, the Radio Intelligence Controller (RIC) of the O-RAN paradigm bestows the network with optimised radio resource allocation, load balancing or energy efficiency functions, among others. Nevertheless, this dependency on data opens new security vulnerabilities, as attackers can alter data properties and steer ML models to underperform or degrade. Conversely, the developed mitigation strategies are effective, but they generate a computational load which, in consequence, results in an energy cost generally overlooked, even in the current energy-awareness context. In this work, consumption of a defence technique is characterised, and the challenges raised by the triad of ML accuracy, robustness and energy efficiency are outlined.

\end{abstract}


\section{Introduction}
\label{sec:introduction}


The integration of Machine Learning (ML) algorithms into cellular networks, particularly through frameworks such as Open Radio Access Networks (O-RAN), has resulted in advancements in communication efficiency, flexibility and delivery of data-driven services \cite{groen2024implementing}, through specialised applications, namely, xApps and rApps. The ML models enable proactive network management with advanced functions accounting for the complex ecosystem created by interactions between users and the network, heterogeneous traffic demand and fluctuating conditions of wireless channels \cite{bib:energy_efficiency_survey}. Despite the potential network enhancement, the cumbersome datasets and dependence on ML algorithms have created new security vulnerabilities beyond the classical cyber-attacks focus on this attack surface \cite{oranSecurity}. While research efforts increasingly address O-RAN security, they predominantly focus on securing interfaces and communication protocols rather than protecting the underlying ML algorithms themselves.

The mentioned gap in focus of attention leaves ML models vulnerable to adversarial attacks that could compromise network decision-making, such as data poisoning, where malicious data is introduced to corrupt models by learning incorrect patterns, or input manipulation, where attackers alter the data inputs to achieve undesired outcomes. The poisoning attack may influence the long-term O-RAN function, because the altered model may not raise any suspicion from the stakeholders and remain in the deployed environment for a prolonged period and even distort the network managers’ planning when their decisions are based on the ML performance. Likewise, the attacker can introduce a backdoor for controlling when an unwanted behaviour is executed, remaining hidden until it is activated. Strategies for filtering malicious data or creating robust ML models have been widely studied \cite{bib:threats_defense_machine_learning_poisoning}. These approaches can demand a high consumption of time and resources.


Along with security, energy efficiency is gaining relevance in network design to incorporate sustainability as one of the goal metrics, because of environmental and economic reasons, while the service performance exigency is also satisfied \cite{bib:energy_efficiency_survey}. The consumption by telecommunication networks has a global impact with a rising tendency, and the Radio Access Network (RAN) segment can contribute to more than $70\%$ of power usage. For example, 5G’s flexibility and adaptability for families of services (such as URLLC, eMBB and mMTC) requires functions such as power control, scheduling, resource allocation, considering the spatial distribution and spectrum sharing for meeting the strict requirements (low-latency, high-reliability, large volume of data, high amount of connected devices, etc.). At its core, energy efficiency in cellular networks is significantly defined by intelligent resource management, which relies heavily on optimisation tasks and forecasting capabilities. For these purposes, Artificial Intelligence (AI) algorithms are also a resort solution. However, apart from infrastructure energy costs, the increasing complexity of AI algorithms has intensified the energy consumption derived from their high computational load \cite{bib:energy_cost_ml}.



A key challenge emerging from the integration of ML into O-RAN is to ensure both security and efficiency within the network, considering its current and future outreaches of these algorithms in network consumption. This trade-off between securing ML models and maintaining energy efficiency highlights the intricate balance that modern cellular networks must achieve. Ensuring robust security without compromising energy-efficient operations is thus a pressing challenge, particularly as networks continue to evolve toward sustainability-oriented and ML-optimised infrastructures, where AI security vulnerabilities could directly hinder energy-saving. Moreover, these security mechanisms often introduce additional computational overhead, increasing energy consumption, which is not typically measured.

This research aims to address the gap between ML security and energy-efficient operations by focusing on the impact of threats and their mitigation on overall system performance. For this purpose, the analysed classification of O-RAN threats will serve as context for the poisoning attack, whose variants and mitigation strategies are described to exhibit the heterogeneity of cases and scenarios. Moreover, the results aim to illustrate the energy cost of a mitigation technique and to provide insights for network design targeting a resilient, efficient, and sustainable cellular network of the future. The formulated discussion is carried out to exhibit some practical considerations of the delicate balance of all parameters.

This paper is structured into seven sections whose content is as follows. In Section \ref{sec:threat_model}, different threats within the O-RAN paradigm are described. Then, the paper in Section \ref{sec:attack_mitigation} focuses on the poison attack, defining the classification of variants and a series of mitigation strategies. The common energy analysis and approaches are defined in Section \ref{sec:model_consumption}. The relationship between mitigation and energy is explored in a real testbed in Section \ref{sec:testbed}, then a discussion is carried out in Section \ref{sec:discussion}. Finally, the conclusions are available in \ref{sec:conclusions}.

\section{O-RAN ML threat model} 
\label{sec:threat_model}

The integration of ML algorithms is crucial to improve the operational efficiency of O-RAN. The classical monolithic architecture from a single vendor is replaced by a composition of new elements, which are interoperable although they are from different vendors: Central Unit (CU), Distributed Unit (DU) and Radio Unit (RU). These components interact via new standardised interfaces and elements in the O-RAN paradigm. Among them, the Non-RT RAN Intelligent Controller (RIC) via rApps is designed to facilitate intelligent RAN optimization by offering policy-based guidance, managing and enriching ML models, and providing information to the Near-RT RIC function. In contrast, the Near-RT RIC can employ ML to deliver near-real-time control and optimisation of RAN elements and resources via data collection and actions over the E2 interface. In this case, the intelligent capabilities are facilitated by the xApps. In both cases, the ML algorithms running in the RIC could be exploited to gain inference and control over the rApps and xApps by  attacks \cite{liyanage2023open}.

The O-RAN Alliance \cite{oranSecurity} has identified critical assets, vulnerabilities, and security requirements within the O-RAN architecture, providing a comprehensive risk assessment framework while leveraging security principles such as secure boot to mitigate AI/ML-driven threats. In \cite{groen2024implementing}, the authors provide a holistic evaluation of O-RAN security across three key dimensions: open interfaces, intelligence, and platforms, highlighting risks from AI-native closed-loop control and cloud-based virtualized functions. Moreover, given the new infrastructure, \cite{groen2024securing} analyses security risks in O-RAN’s open interfaces, in which the impact of encryption mechanisms on system latency and throughput is considered. Also, centred around one O-RAN element, in \cite{chiejina2024system}, the developed malicious xApp executes adversarial attacks on ML-based interference classification xApps in the near-RT RIC to evaluate the impact of these attacks on network performance.

As ML-driven intelligence becomes more extensive, safeguarding these models against adversarial threats is important to maintaining reliable network performance. In this sense, the Open Worldwide Application Security Project (OWASP) has classified the following security attacks as the top 10 \cite{OWASP_Top10}: input manipulation, data poisoning, model inversion, membership inference, model stealing, AI Supply chain, transfer learning, model skewing, output integrity, and model poisoning. These attacks impact the security aspects of the three vertexes of the Confidentiality/Integrity/Availability (CIA) triad, and different phases of the ML lifecycle, such as the training phase, the inference phase, the deployment, and the upgrades to optimise the models.

Attacks can be classified into three main categories based on the knowledge of the attacker with respect to the ML algorithms deployed in the network \cite{O-RAN-WG11-AI-ML-security}:

\begin{itemize}
    \item Black-box: the attacker can only access the ML model and its corresponding output. Despite its limitations, the real model can be approximated and, then, attacks can be designed and transferred to the actual case.

    \item Grey-box: the attacker gains partial knowledge of the model or the parameters.

    \item White-box: the attacker has obtained full knowledge about the ML models, along with their hyperparameters. In controlled scenarios, this explores the upper limit of the attacker's damage to the network.
    
\end{itemize}

Despite the great effort from the O-RAN Alliance in defining vulnerabilities \cite{O-RAN-WG11-AI-ML-security}, the current status of available solutions is partial and incomplete \cite{bib:xApp_store}, regarding the implemented interfaces which satisfy all the O-RAN requirements and the capacity of testing the infrastructure.

\section{Altering attacks} 
\label{sec:attack_mitigation}

The focus of this paper is on data poisoning attacks due to the huge amount of processed data in the O-RAN infrastructure, enabling the introduction of altered samples in the datasets. Furthermore, their impact on O-RAN can dilute the target RIC function  performance with a long-lasting effect.

\subsection{Poisoning attacks} \label{DP}

Poisoning \cite{bib:threats_defense_machine_learning_poisoning} is a type of attack where the objective is to alter the model's behaviour; for example, this manipulation can result in developing backdoors that could be exploited or directly to degrade the normal behaviour of the ML model. Such attacks on the cellular network have the potential to induce the contamination of normal samples, which could generate false feedback, the manipulation of private data, or a misleading result on the detection system, thereby increasing the probability of receiving malicious traffic \cite{sun2020machine}.

\begin{figure*}[htb]
\centering
  \includegraphics[clip,width=1.65\columnwidth]{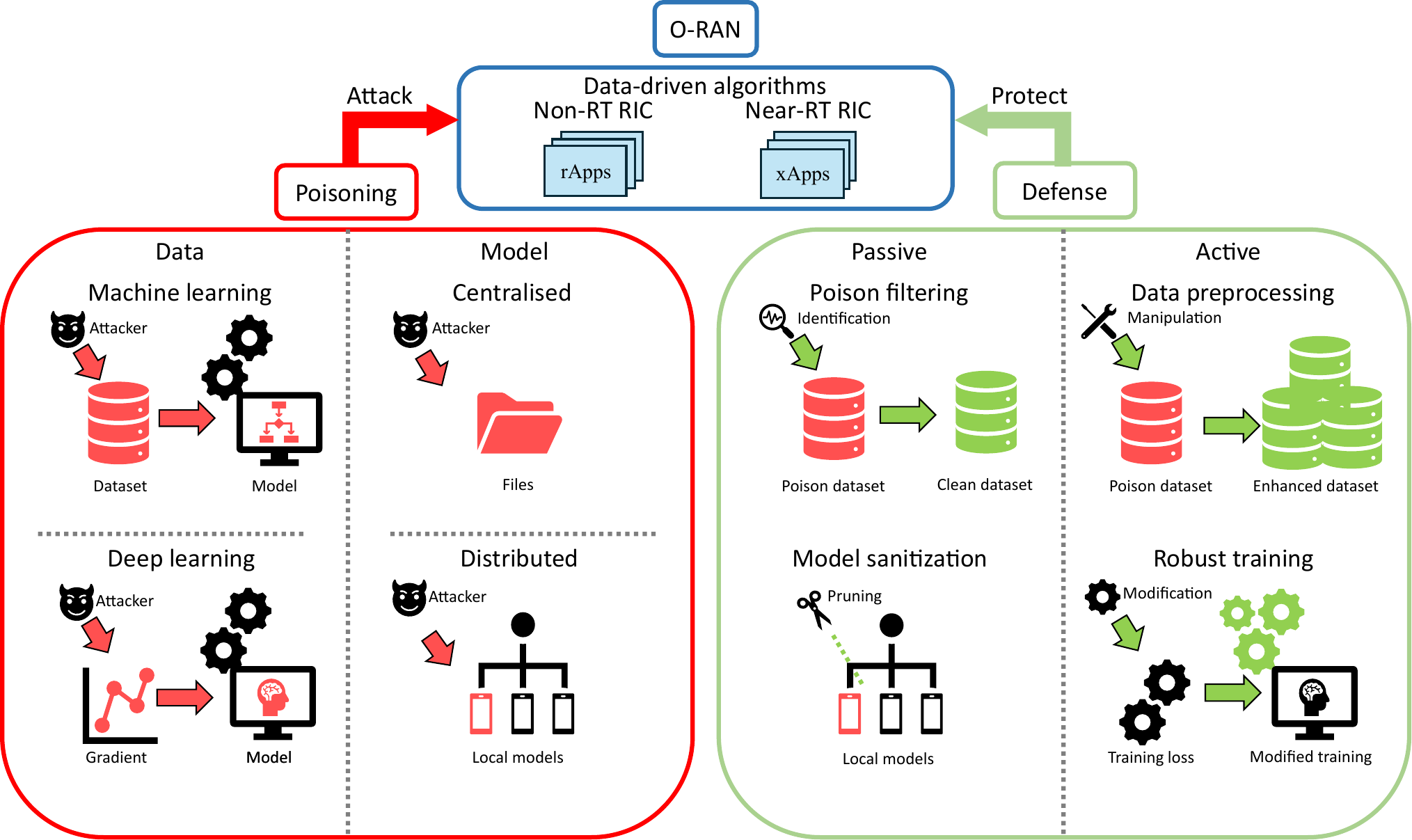}
\caption{ML data poisoning attacks in O-RAN.}
\label{fig:attack_mitigation}

\end{figure*}

Depending on the attacker's goal, poisoning attacks can be classified into the following categories, which are represented in Figure \ref{fig:attack_mitigation}:

\begin{itemize}
    \item \textit{Data}. The basis of ML consists of training a model with proper data for predicting a target label with the input features. Consequently, an attacker can alter the relationship between input and output through manipulations of the data. This is further categorised in the following classes:

    1) \textit{Machine Learning}. Traditional algorithms are designed with heterogeneous methodologies, such as support vector machine, random forest, logistic regression, etc. Therefore, the attacker tries to modify the training samples according to different approaches. It can target deliberately mislabel the samples to induce an incorrect training. The process can be designed to maximise the effectiveness of modifying data, in order to search for a global optimal under certain constraints. Moreover, the attacker can operate under a budget to limit the alteration to craft poison samples or noise to modify the already existing ones.

    2) \textit{Deep learning}. A neural network can be stacked with a substantial number of hidden layers whose elements’ weights are optimised through a gradient-based algorithm. For this reason, a vast part of the literature centre its attention on these algorithms. In this case, the attacker can approximate the original algorithm behaviour in order to iteratively perturb the samples to a set adversarial target or use a strategy to choose the most influential samples. If not, the attacker can also create generative models to craft poison samples. Otherwise, the approach can keep the original data but slightly introduce non-noticeable alterations, which, even in white-box scenarios, spoil the ML model.

    \item \textit{Model}. Instead of the training samples, the attacker can exploit the models' files to alter their behaviour. This is less explored in the literature, and it can be further classified into:

    1) \textit{Centralised}. The algorithm, with its parameters, is stored in files which can be fine-tuned to alter its outputs. 

    2) \textit{Distributed}. An apparently benign intruder acts within the model update process to poison the global model.

\end{itemize}





\subsection{Defense strategies}
\label{sec:defense_strategies}

Defensive strategies against data poisoning can take action in different stages, from reducing the damage already induced by attackers to preventing the root causes from influencing the models. Usually, the developed technique cannot avoid the attack effect completely, although it minimises it at the cost of the model’s accuracy. An O-RAN example can be found in \cite{chiejina2024system}, where the authors demonstrate the effectiveness of a distillation method for mitigating the attack with an accuracy over $90 \%$.

According to \cite{bib:threats_defense_machine_learning_poisoning}, the countermeasures’ taxonomy can be as follows, as illustrated in Figure \ref{fig:attack_mitigation}:

\begin{itemize}
    \item Passive. It mitigates the attack effect and influence through sanitisation by relabelling or removing malicious elements.

      1) \textit{Poison filter}. It identifies poisoned data to remove or relabel them. This can be based on outlier removal samples from class centroids or on the features' incosistency to neighbour samples. Other approaches train several base models on disjoint datasets, made from the original to compare the results and remove the diverging ones. Moreover, in the case of owning a clean dataset, it can be used as a validation set for comparing unclear samples.

      2) \textit{Model sanitization}. In a centralised model, the objective is to reduce the influence of contaminated parts, such as pruning neurons in a neural network. In the case of distributed learning, the updates of the global model is inspected to compare the characteristics of benign users to suspicious ones.

    \item Active. These approaches act before the threat has poisoned, analysing the data or preparing the model to be resilient against it. 

     1) \textit{Data preprocessing}. It consists of cleaning, enhancing and transforming the dataset to reinforce the model. The current trend prefers leaning toward adding samples with a data augmentation strategy to improve the model generalisation.
            
     2)  \textit{Robust training}. It creates a model which can counter anomalous training inputs. For this purpose, modifications of the model loss calculations, noise addition to samples, or ensemble methods to distribute the poison influence are applied.

\end{itemize}

The computational load involved in these mitigations is usually ignored or considered exclusively in time variables, such as the latency introduced in the system. However, the pertinent energy consumption can be significant due to the same computational expenses, although is not explored to the best of the authors’ knowledge.

\section{Model consumption}
\label{sec:model_consumption}

To understand the energy impact of the mitigation strategies at the system level, this section will describe, in the first place, the context of consumption. From this information, the relevance of ML in its current status is explored to provide actionable insights, easing its integration into practical real-world deployments.

\subsection{Energy saving}

In cellular networks, energy expenses are examined from the contributions of each network component \cite{bib:energy_survey}, defining analytical expressions for this purpose. In further detail, the physical aspects are considered, such as transmitted power, MIMO configuration, analogue and digital components, as well as the virtual elements, such as O-RAN functions.

The wireless network environment is complex and stochastic by nature; the relationship between  traffic requirements, user mobility, interference, and channel variations in time and space make system-wide optimisation a hard problem. From 3GPP and ETSI, energy related metrics are proposed, which associate consumption with network specifications, such as data volume, coverage or latency. These metrics enable the evaluation of the adequacy of solutions whose objective is to increase the system energy efficiency. For example, the literature proposes methodologies for deactivating network elements or reducing the radio resource waste. 

\subsection{ML role}

Given the interactions of all energy variables and the consequent challenges for reaching efficiency, AI/ML literature proposes it as an enabler for decreasing unnecessary consumption \cite{bib:energy_survey}. In this regard, data-driven algorithms analyse network parameters to forecast traffic or energy demand, opening up the possibility for proactive efficiency strategies. Nevertheless, the ML computational load has been steadily increasing more than 300,000 times with respect to previous years, and this is expected to be higher due to the popularisation of LLMs or similarly complex models \cite{bib:energy_cost_ml}. The data collection, processing and storage, together with the cost of training algorithms, are significant factors in increasing energy cost.

Likewise, ML defence implies a high computational load and processing, as the mitigation is directly related to data manipulation and generation of complementary ML models. However, energy and security are not contemplated together; they are examined as independent dimensions of network design. Solely, the study of ML models focuses on improving their accuracy or developing more efficient algorithms within the green AI paradigm, applied to network functions \cite{bib:energy_survey}. Hence, this paper fills the gap, offering empirical measurements of the energy cost of mitigating a poisoning attack.

\section{Mitigation performance and evaluation}
\label{sec:testbed}

In this section, empirical experiments are conducted to understand the consumption of a mitigation technique for a poisoning attack. For that purpose, in the first place, the attack actors in the scenario are described. Then, the next subsection presents the poisoning and its mitigation. Finally, the energy cost is measured and analysed.

\subsection{Testbed}

\begin{figure*}[htb]
    \centering
    \includegraphics[width=2\columnwidth]{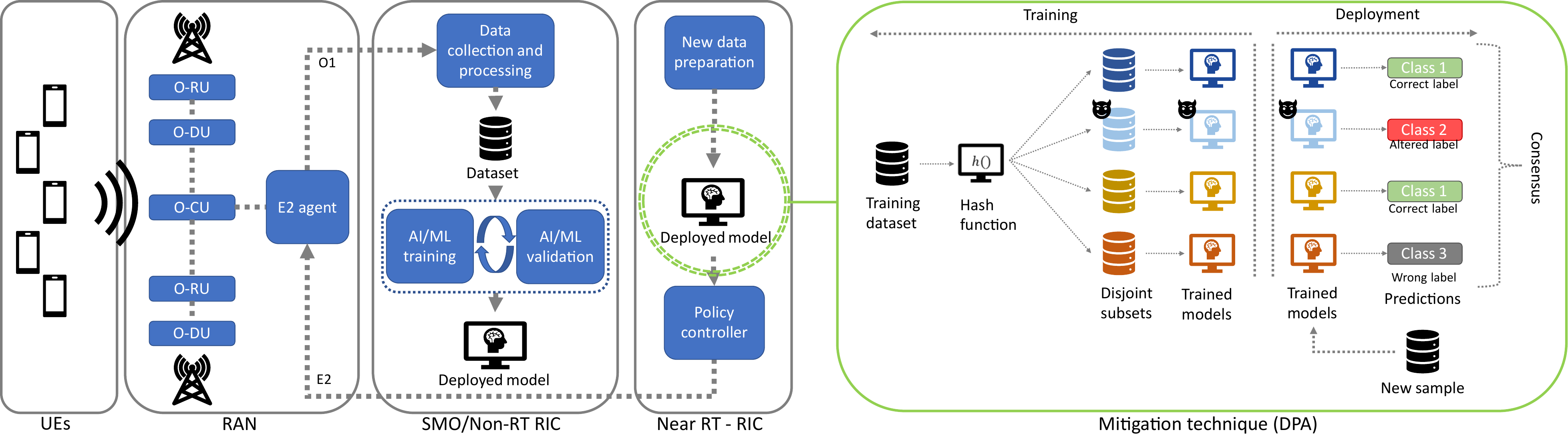}
    \caption{Context of poisoning in O-RAN and the chosen mitigation technique.}
    \label{fig:experimental_cases}
\end{figure*}

In O-RAN, the Near-RT RIC provides a quick response to environmental changes and adapts the network to other conditions, and within this element, previously trained models are deployed as represented in Figure \ref{fig:experimental_cases}. Nevertheless, the O-RAN paradigm is an ongoing development with incomplete parts \cite{bib:xApp_store}. Thus, an attacker can access the interfaces to inject or modify training samples, creating a need for a mitigation technique to avoid network degradations, although this generates energy consumption. Here, a proof of concept is carried out with the model ResNet-32, evaluated under different conditions: the normal case, where the attacker has no influence; the poisoned case, with altered data; and the mitigated case, where the attack bias has been reduced. The defence performance in terms of data protection capabilities from poisoning attacks will be examined. Moreover, the time and power (in watts) consumed during the execution of the security tests will be measured.

The widely used CIFAR-10 dataset has been chosen as the target to provide generalisable results to compare with the literature. This dataset contains 10 classes (6000 images per class), with a total of 60000 images (split into 50000 for training and 10000  for testing). Two evaluation metrics are used: \textit{accuracy}, i.e. the ratio of correct labels compared to the evaluated ones, and \textit{target misclassification}, which is the rate of source data correctly classified in a clean dataset and misclassified in the poisoned one \cite{DEF-2023-Incompatibility_Clustering_as_Poisoning_Defense}. Higher accuracy and lower targeted misclassification are desirable.

The poison attack uses the implementation from \cite{DEF-2023-Incompatibility_Clustering_as_Poisoning_Defense} to alter the data and to perturb the prediction algorithm, generating misclassifications. The poisoning rate in the dataset will range from 0.05 to 0.40 in steps of 0.05 to evaluate how the defensive technique behaves against different levels of poisoning. The maximum poisoning has been limited to 0.40 because, beyond this point, the results start to decrease rapidly as the number of poisoned data reaches about half.

On the other hand, the Deep Partition Aggregation (DPA) method, represented in Figure \ref{fig:experimental_cases} and is adapted from \cite{levine2021deep}, has been chosen as the mitigation technique due to its known effectiveness \cite{bib:threats_defense_machine_learning_poisoning}. It consists of splitting the dataset into \textit{k} partitions and applying a hash function to each sample to determine which partition it is associated with. The result of the hash function must depend only on the sample value and a model is trained for each partition. The final output is the consensus of all the models.

Two experimental testbed setups are used due to equipment availability. The first one is a PC (PC1) equipped with a 13th Gen Intel® Core™ i9-13900K CPU. It has 64GiB of DDR4 RAM, a 512GiB NVMe drive, and Intel UHD Graphics 770. The second one (PC2) is equipped with a 13th Gen Intel® Core™ i9-13900K CPU. It has 64GiB of DDR4 RAM, a 512GiB NVMe drive, and Intel UHD Graphics 770. Both are running Ubuntu 22.04.5 LTS. The power recollection has been performed using Scaphandre, a Linux tool for providing energy consumption metrics for individual processes. Focusing on the energy metrics,  \textit{scaph\_process\_power\_consumption\_microwatts}, which allow for measuring power consumption in microwatts for individual processes.

\subsection{Mitigation performance}
\label{sec:mitigation_performance}

\begin{figure}[htb]
    \centering
    \includegraphics[width=1\linewidth]{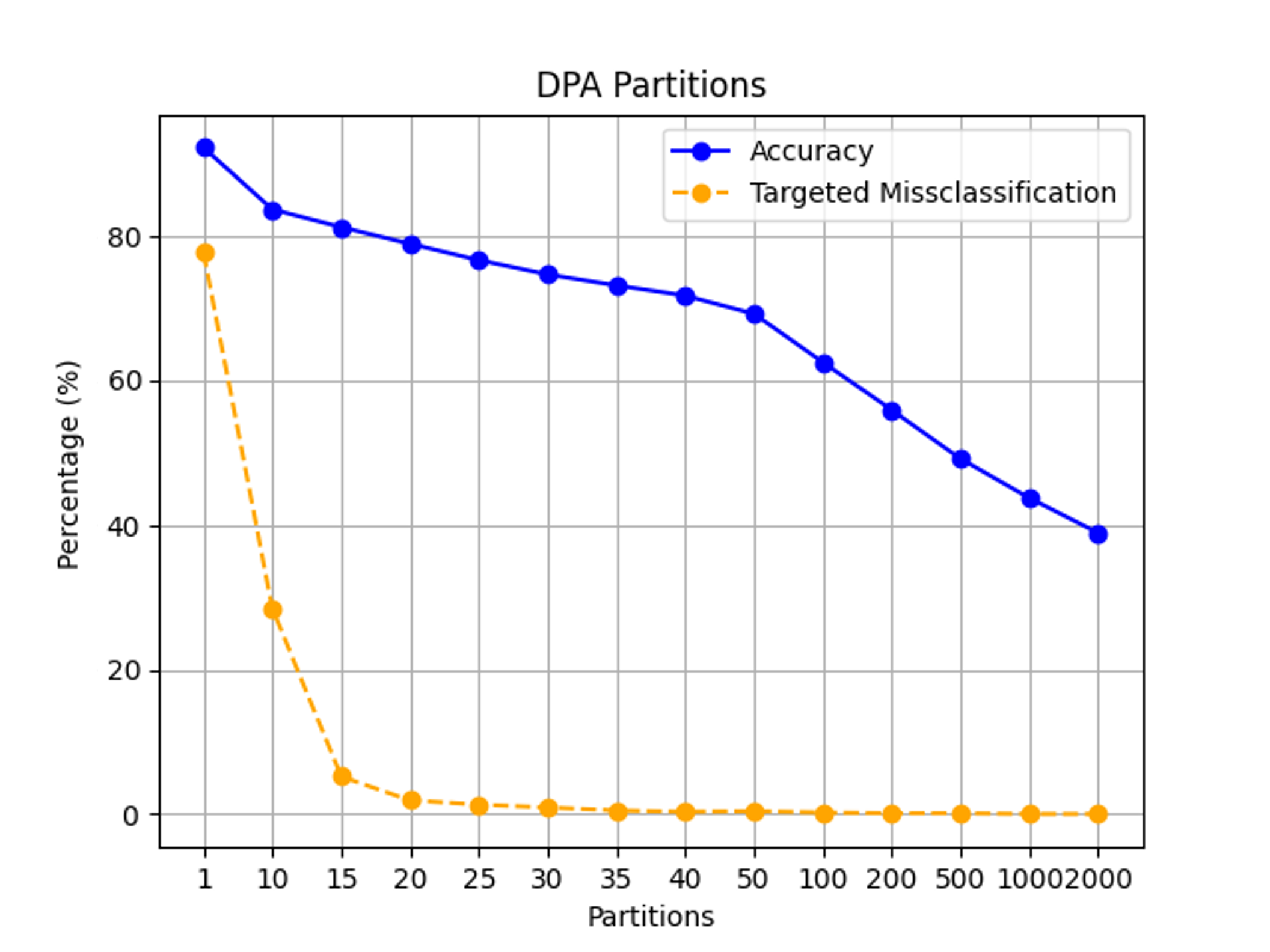}
    \caption{DPA performance using different numbers of partitions with a constant poisoning rate of 0.2.}
    \label{fig:test-dpa-partititons}
\end{figure}

Regarding the DPA technique, before proceeding with the normal test execution, it needs to determine the optimal number of partitions $k$ by finding a balance between system accuracy and targeted misclassification. For this purpose, a partition sweep is evaluated with a 0.2 poisoning ratio, considering this representative of a realistic setting with a limited presence of altered samples. Results for this test are shown in Figure \ref{fig:test-dpa-partititons}.

As seen, increasing the number of partitions decreases \textit{accuracy}, while \textit{targeted misclassification} improves (i.e. it diminishes). This is because data is divided across different partitions, so each partition has less information to predict accurately, but the combined results are more robust. However, a good \textit{targeted misclassification} is not significant when the accuracy is deficient, as the results of the first metric depend on the second one. Thus, 15 partitions are chosen as the optimal DPA configuration. Although the \textit{targeted misclassification} is not zero, the accuracy remains above 0.8, ensuring balanced results.

\begin{figure}[htb]
    \centering
    \includegraphics[width=\linewidth]{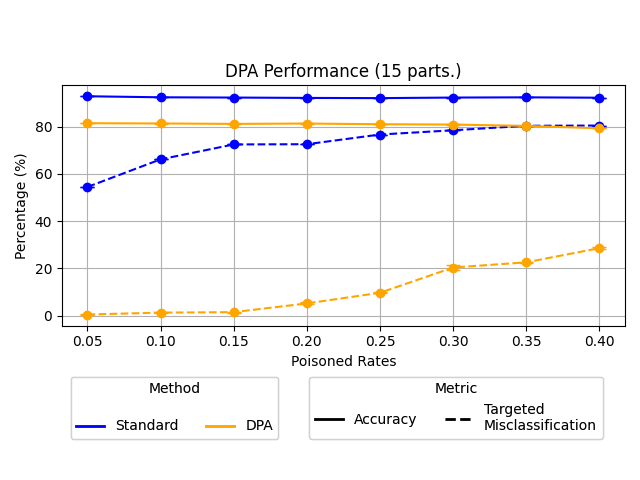}
    \caption{Standard (baseline) and DPA performance using 15 partitions poisoning data from 0.05 to 0.40.}
    \label{fig:dpa_poisoning_performance}
\end{figure}

Once the optimal number of partitions is determined, the performance tests for DPA are carried out and represented in Figure \ref{fig:dpa_poisoning_performance}. In the first place, \textit{accuracy} reveals that integrating DPA protect the model, preserving decent performance around $80\%$ in all cases, demonstrating $k = 15$ validity in mitigation. Conversely, it can be observed that increasing the poisoning rate leads to a higher \textit{target misclassification}, while the system \textit{accuracy} presents a lower slope. With a poisoning rate $\leq 0.15$, the \textit{target misclassification} value remains $<1\%$. However, since a 0.2 poisoning rate, a higher increase in the \textit{target misclassification} is observed at each step, reaching a value of $28\%$ at a poisoning of 0.4. In the last cases, DPA cannot reduce the contaminated situations as the number of partitions is not enough to mitigate them.

\subsection{Energy consumption} 
\label{xAppR}

During DPA execution, the watts consumed and the time period are recorded, enabling the calculation of energy consumption in Watt-Hours (Wh).

\begin{figure}[htb]
    \centering
    \includegraphics[width=0.8\linewidth]{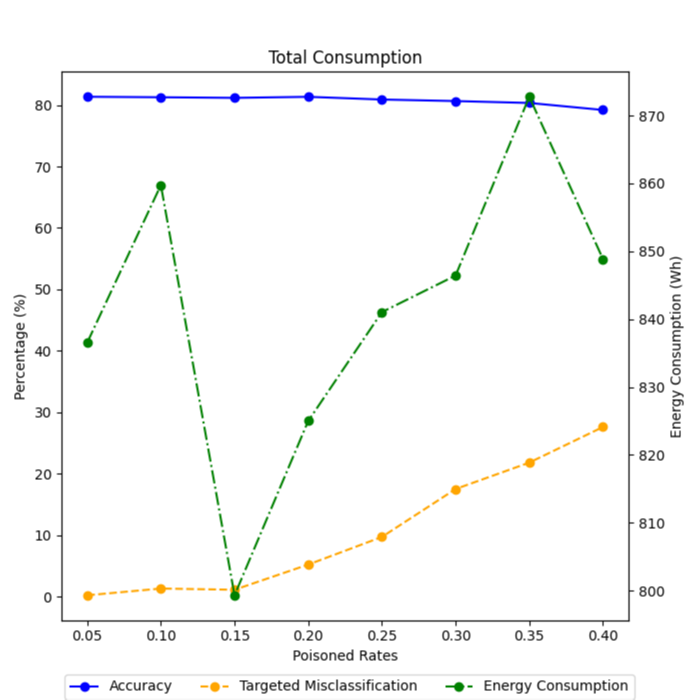}
    \caption{DPA energy consumption with different poisoning rate.}
    \label{fig:dpa_poisoning_energy}
\end{figure}

In the first place, the system is evaluated in PC2 with an increasing poison rate to examine its influence. The corresponding results are depicted in Figure \ref{fig:dpa_poisoning_energy}. Although there is a growth after 0.15 of poisoning, no clear pattern is denoted in energy consumption; hence, poisoning rate is an independent variable from energy waste. This is expected; the differences in data features due to the attacks do not affect DPA analysis, and they are processed as normal samples.

\begin{figure}[htb]
    \centering
    \includegraphics[width=0.8\linewidth]{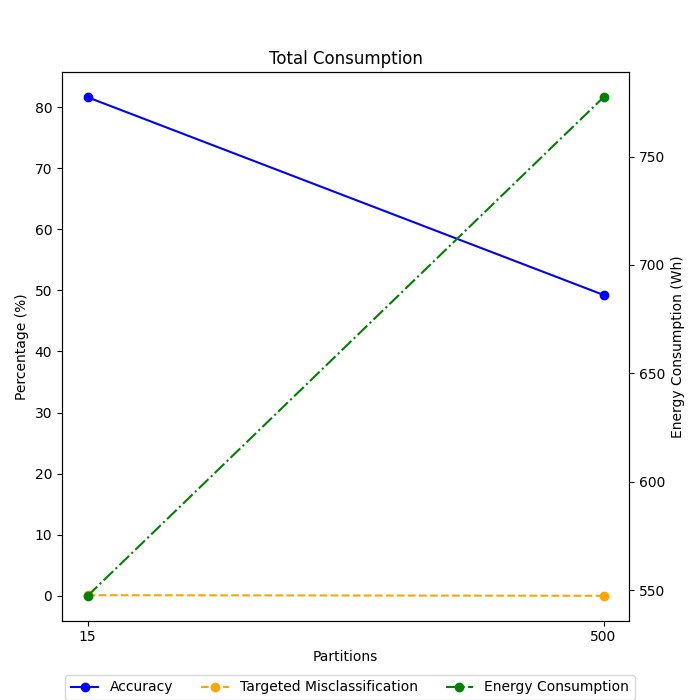}
    \caption{DPA energy consumption with different $k$ partition configurations, with 0.05 of poisoning rate.}
    \label{fig:dpa_partition_energy}
\end{figure}

Conversely, the DPA configuration can alter its generated computational load, so $k$ is modified in PC1. The energy consumption with different numbers of partitions is analysed to determine if increasing the number of models to train leads to higher energy consumption. The previously chosen case $k = 15$ is compared to $k = 500$, when the \textit{accuracy} is approximately $50\%$. Because of infrastructure availability, only these cases are examined. In both cases, the poisoning rate is 0.05, ensuring the presence of poisoned samples for DPA. The results are exhibited in Figure \ref{fig:dpa_partition_energy}, where the execution with 15 partitions consumes significantly less energy than with 500 partitions, with a difference of more than 200 Wh. This is an increase over $10\%$ of algorithm consumption because of the mitigation configuration. This could be explained by the fact that the system needs to train a total of 500 partitions from scratch, which, even though they contain less data, results in higher overall consumption compared to training a smaller number of partitions with their respective models.



\section{Discussion} 
\label{sec:discussion}

\subsection{Practical constraints}

The choice of ML models for network functions must contemplate the entire ML lifecycle from data gathering to model inference and practical facets such as data processing and computational implications. Likewise, energy resource consumption must be a new dimension in network design. When AI defense algorithms are included in the RIC (or any data-driven component), it can entail training surrogate or complementary models that consume a noteworthy amount of computational resources  \cite{bib:energy_cost_ml}, and this translates into an energetic cost. Among the defense categories described in Section \ref{sec:defense_strategies}, certain methods to enable robust training or data pre-processing may create models which are less moldable to an attacker and not require extra protection. In this regard, lower ML sensitivity to perturbations in samples’ features can be a natural defence against attackers and a guideline to outline a plan for countering attacks without sacrificing energy efficiency.

Moreover, although it is not studied here, as it is outside of this paper’s scope, O-RAN network elements must satisfy time restrictions, which are difficult to meet for mitigation methods. For example, the DPA solution requires training multiple independent models, each one with its own training and inference time. In this sense, suitable quick response mitigation techniques are scarce.

\subsection{Future paths}

With respect to the ML performance metrics, general accuracy or similar parameters (recall, F1, etc.) describe a shallow adequacy of the ML model in the AI security context. Dedicated indicators, beyond the \textit{targeted misclassification}, can depict a more detailed impact of the mitigation technique. Moreover, this should integrate energy costs to quantify the consumption price of potential improvements in security.

On the other hand, the energy characterisation of any mitigation technique is a double-edged sword as it enables energy saving, but it also opens up new breaches for attacks. The unique consumption pattern of defence may reveal specific underlying processes in the system. Hence, an attacker can switch to an offensive approach for which the cellular network is less prepared, or simply alternating methodology to find the attack to maximise the network consumption. For this reason, before a network is planned and deployed, its architectural design should integrate masking methods to hide energy consumption.

\section{Conclusions} 
\label{sec:conclusions}

Data-driven algorithms are growing in their implementation within cellular networks, from low-level radio optimisation to high-level resource management. Despite AI's potential, vulnerabilities arise due to its reliance on data, which can be altered by attackers to lead the model to deficient performance. While countermeasures are developed, their success involves energetic expenses. In this work, this has been characterised for DPA, and demonstrates that the trade-off in the triad composed by the model accuracy, defence capacity and energy cost must be considered during the network design.


\section*{Acknowledgments}

This work has been partially funded by Ministerio de Asuntos Económicos y Transformación Digital and European Union — NextGenerationEU within the framework “Recuperación, Transformación y Resiliencia y el Mecanismo de Recuperación y Resiliencia” (MAORI). Funding for open access charge: Universidad de Málaga/CBUA.


\bibliographystyle{IEEEtran}

%

\section*{Biography}

\vskip -2\baselineskip plus -1fil

\begin{IEEEbiographynophoto}{ADRIÁN LOSADA CASADO (adrian.losadac@alumnos.upm.es)} received the Computer Science and Engineering in 2019 and Master in Computer Science and Computing in 2020, both from University Carlos III (Madrid). His experience covers several programming languages and applying AI. 
\end{IEEEbiographynophoto}

\vskip -2\baselineskip plus -1fil

\begin{IEEEbiographynophoto}{HAO QIANG LUO-CHEN (hao@uma.es)}
is a Ph.D. candidate in Telecommunications Engineering from the University of Málaga, Spain. His research interests are self-healing, user localisation, IA security and energy efficiency.
\end{IEEEbiographynophoto}

\vskip -2\baselineskip plus -1fil

\begin{IEEEbiographynophoto}{DAVID SEGURA (dsr@ic.uma.es) } received his Ph.D. degree from University of Málaga (2024), with a focus on the performance assessment of cellular networks applied to the industrial scenario. 
\end{IEEEbiographynophoto}

\vskip -2\baselineskip plus -1fil

\begin{IEEEbiographynophoto}{CARLOS SIMÓN ÁLVAREZ-MERINO (csam@uma.es) } received the Ph.D. degree in 2024 on the topic of Indoor Localisation.
\end{IEEEbiographynophoto}

\vskip -2\baselineskip plus -1fil

\begin{IEEEbiographynophoto}{MILAN GROSHEV (mgroshev@faculty.ie.edu)} received the Ph.D. degree in telematics engineering from the University Carlos III Madrid in 2022. 
\end{IEEEbiographynophoto}

\vskip -2\baselineskip plus -1fil

\begin{IEEEbiographynophoto}{EMIL J. KHATIB (emil@uma.es) } (Member, IEEE) received the Ph.D. degree in 2017. He is currently an Assistant Professor (University of Málaga), researching security and localization in B5G/6G networks.
\end{IEEEbiographynophoto}

\vskip -2\baselineskip plus -1fil

\begin{IEEEbiographynophoto}{RAQUEL BARCO (rbarco@uma.es) } is Ph.D. in telecommunications from the University of Málaga, where she is currently Full Professor, specialized in mobile communication networks and smart-cities.
\end{IEEEbiographynophoto}

\end{document}